\documentclass[conference]{IEEEtran}
\usepackage{cite}
\usepackage{amsmath,amssymb,amsfonts}
\usepackage{algorithmic}
\usepackage{graphicx}
\usepackage{textcomp}
\usepackage{xcolor}
\def\BibTeX{{\rm B\kern-.05em{\sc i\kern-.025em b}\kern-.08em
    T\kern-.1667em\lower.7ex\hbox{E}\kern-.125emX}}
\begin{document}

\title{Food waste estimation using Received Signal Strength Indicator\\}

\author{\IEEEauthorblockN{Daniel Koohmarey}
\IEEEauthorblockA{\textit{Computer Science and Engineering } \\
\textit{University of California, San Diego}\\
San Diego, USA \\
dkoohmar@eng.ucsd.edu}
\and
\IEEEauthorblockN{Nitish Nagesh}
\IEEEauthorblockA{\textit{Electrical and Computer Engineering} \\
\textit{Technical University of Munich}\\
Munich, Germany \\
nitish.nagesh@tum.de}

}

\maketitle

\begin{abstract}
Food loss and waste is a problem with rising global recognition. Attempts to combat food waste at institutional levels such as college campus dining halls require insight into food waste statistics to evaluate efforts to reduce food waste. Gathering such data is currently costly, requiring manual sorting and weighing of trash bin contents. We propose an alternate method of estimating the food waste in trash bins using a radio frequency (RF) transceiver’s Received Signal Strength Indicator (RSSI) to estimate the attenuation caused by the trash contents. We propose food attenuates the radio frequency signal at greater rates than the paper and plastic trash present, allowing us to correlate and estimate the amount (weight) of food waste present with the observed RSSI values. We witness promising results while testing this approach.
\end{abstract}


\section{Introduction}

Food loss and waste is a problem with rising global recognition as evident by the 2030 Agenda for Sustainable Development that aims to halve per-capita global food waste by 2030\cite{fao_1}. Globally, the U.N Food and Agriculture Organization estimates one third of food for human consumption is lost or wasted\cite{dusseldorfsave}. The actual impact of food loss and waste goes beyond just the material lost, but also in the energy and resources it takes to grow the food and the related related emissions and climate impact. Food waste accounts for roughly 8\% of greenhouse gas emissions, according to an additional study by the United Nations Food and Agriculture Organization\cite{fao_2}. Restaurants and food service providers handle large volumes of food, and are important targets to adopting changes to reduce food waste for their own environmental and financial benefit.

At the University of California San Diego (UCSD) Price Center dining hall, metrics with regards to food waste are not easily gathered. Costly one time manual analysis of trash contents to determine food waste requires manual sorting and weighing of trash which fail to capture any temporal trends (if the waste profile changes over time). The UCSD dining community is interested in reducing food waste and currently uses a logistical tracking system known as FoodPro by Aurora to estimate current food waste at the inventory level\cite{ucsd_zerowaste}, but not the food waste from what is served. Data on the food wasted that is actually served would allow the dining hall to develop strategies to further minimize food waste.

Approaches to approximate food waste given contents of waste bins is challenging due to the heterogeneous nature of trash, it is traditionally hard to determine the amount of food waste without sorting. The closed, dirty nature of the trash cans and disposable boxes also limit the effectiveness of vision processing based solutions. Additionally the adoption of any new technology to address the waste issue in trash cans would be an additional financial burden on the institution that would need to be minimal to be viable. An existing solution would require a sensor that differentiates material or a model to estimate waste built from localized waste data. Existing tools measure whether the bin is full or not, but make no estimation to the amount of food. Food quality estimation techniques exist to determine if food has spoiled to avoid waste\cite{ha2018learning}, but not techniques to estimate the food waste.There is a need for a low cost estimation technique to provide metrics on food waste for those seeking the data to manage their food waste.

We implement an RF based sensing system that relies on RSSI to estimate the level of food waste in a trash bin despite its heterogeneous nature. This is achieved by estimating food waste based on RF attenuation, since RF attenuation changes depending on the material the energy travels through\cite{rappaport1996wireless}. We expect food content to also have higher moisture than other types of paper or plastic waste, which in turn will result in greater attenuation because of the presence of water which is conductive\cite{ryan2003radio}. Food waste should noticeably attenuate the RF signal proportionate to the amount of food waste present at greater levels than other paper or plastic waste. The RF approach allows for the passive “hands off’ estimation of food waste without sorting or modifying the functionality or structure  of existing waste bins. In this way existing waste bin systems may also be retrofitted with sensors and food waste data collected.

\section{Related Work}
A variety of research works related to food waste exist, however they do not address the amount of food waste itself in a trash bin of mixed content. Studies explore the use of RFIDs to evaluate the physical properties of food to determine spoilage, in an effort to save food that may be mistakenly wasted. Such studies evaluate the food quality of beef\cite{nguyen2015wireless}, alcohol\cite{ha2018learning} or milk and dairy\cite{potyrailo2012battery}. There is previous work also done to detect when a garbage bin is full using Zigbee\cite{smart_waste_management}\cite{lata2016iot}, UltraSound and Long Range (LoRa)\cite{cerchecci2018low} but they do not detect the contents of the garbage itself. Logistical systems that track food production\cite{zou2014radio} may also quantify food produced but not wasted by the consumer. At a macro level studies also exist that track food waste at the national level, such as the United States\cite{hall2009progressive}. While these studies are good at understanding the greater global impact of the waste issue, they do not provide any information with local contexts to facilitate local changes. Overall existing research has specific applications in the food waste space but does not track the actual waste present in a trash bin, a metric we consider useful in combating food waste. The use of RSSI as a sensing metric has previously been used to estimate distance\cite{benkic2008using}, but not to estimate attenuation to infer material composition.

\begin{figure}[htbp]
\begin{center}
\centerline{\includegraphics[width=0.48\textwidth]{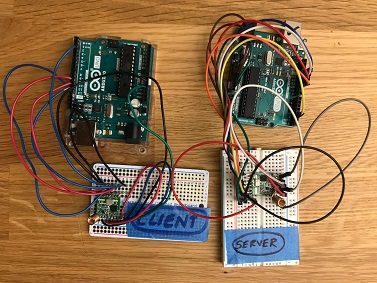}}
\caption{Hardware Configuration.}
\label{hardware_config}
\end{center}
\vspace{-1mm}
\end{figure}

\section{Methodology}

We designed an RF based system for estimating food waste using Arduino Uno's with RFM$22$B-S$2$ SMD wireless transceivers. The transceivers were located on the bottom and top of the trash bin whose food waste was being measured such that the RF signal attenuates through the trash between them. the Received Signal Strength Indicator (RSSI) metric supported by RF transceivers was used to determine the effect of attenuation by measuring the change in strength as a result of change in material between the transceivers. We read the digital RSSI values from the RF transceiver and display the value on a serial monitor to record the results. We transmit multiple test messages from one transceiver to generate and record multiple RSSI values on the receiving transceiver. Throughout our testing we aimed to record a minimum of ten RSSI values to evaluate the stability of the results. The amount of food waste (weight) was estimated with a model developed from the RSSI values recorded when known quantities of food waste existed between the two transceivers. Initial calibration was done to ensure hardware settings were chosen to yield stable RSSI readings. We account for the path loss between the two transceivers by ensuring we get RSSI readings when an empty trash bin is present to contextualize subsequent results. We expected to observe reduced RSSI readings when more food waste existed between the transceivers as food waste should noticeably attenuate the RF signal proportionate to the amount of waste present. We tested our food waste estimation system against actual waste found at the UCSD Price Center dining hall and fruit/vegetable waste from a local grocery store. Finally, we test the accuracy of a food waste estimation model built from the grocery waste data.

\subsection*{Hardware}
We used RFM22B-S2 SMD wireless transceivers for the transmitter and receiver because the transceiver supported a digital RSSI value\cite{electronic2016rfm22b}, allowing us to read the RSSI from a register without any additional measurement instrumentation. For the transceiver antennas we use a simple spring antenna designed for use at 915 MHz with 2.15dBi of gain and 50$\Omega$ impedance (required by the RFM22B). We use the Arduino Uno as it can interface with the RFM22B via SPI and it supports control of the RFM22B using the RadioHead library\cite{mccauley2014radiohead}. Our hardware setup with transmitter (client) and receiver (server) is shown in Fig.~\ref{hardware_config}.

\subsubsection*{Hardware Configuration}
\begin{figure}[htbp]
\begin{center}
\centerline{\includegraphics[width=0.48\textwidth]{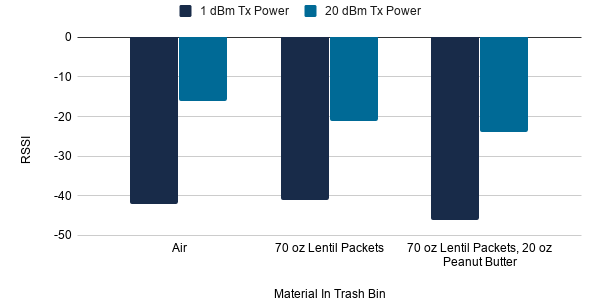}}
\caption{RSSI for 1 dBm and 20 dBm transmitter power.}
\label{rssi_power}
\end{center}
\vspace{-1mm}
\end{figure}

\begin{figure}[htbp]
\begin{center}
\centerline{\includegraphics[width=0.48\textwidth]{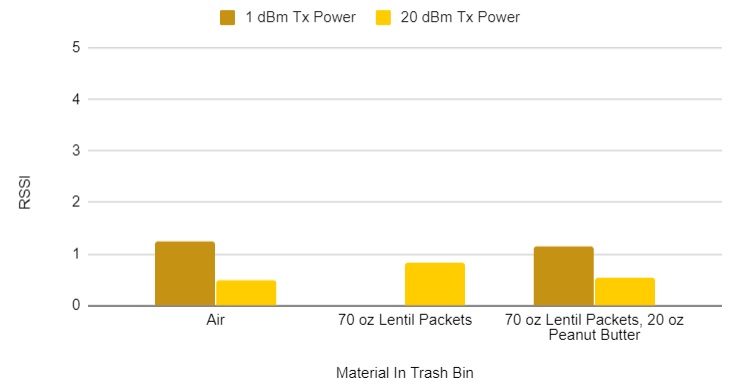}}
\caption{Transmitter Power Standard Deviation.}
\label{std_dev_power}
\end{center}
\vspace{-1mm}
\end{figure}

Before gathering test data with our setup, we needed to finalize configuration options available to us with the goal of achieving maximum stability in future RSSI readings. The RFM22B supported transmitting at powers ranging from 1-20 dBm and frequency from 433-915 MHz. We chose to use 20 dBm transmit power because the calculated path loss at 5 feet (what we considered the average height of a trash bin would be) was 35dB of attenuation\cite{rf_cafe} assuming a -5 dB system gain. The resulting expected RSSI as a result of path loss with 20 dBm transmitter power was -16dB. Evaluating our power level decision in Fig.~\ref{rssi_power}, we observe that the median RSSI differs by roughly the difference in transmitter power which is expected behaviour. A comparison of the resulting standard deviations in Fig.~\ref{std_dev_power} between the 1 dBm and 20 dBm transmitter power further demonstrates that both are stable with low standard deviation. To ensure future readings did not encounter any issues with trash contents causing too much attenuation and inhibiting our signal from being received, we opted to preemptively use the maximum transmitter power available (20 dBm).
For the frequency, because we had chosen an antenna designed for use at 915 MHz, we ensured we configured our transceivers frequency to 915MHz.

\begin{figure}[htbp]
\begin{center}
\centerline{\includegraphics[width=0.48\textwidth]{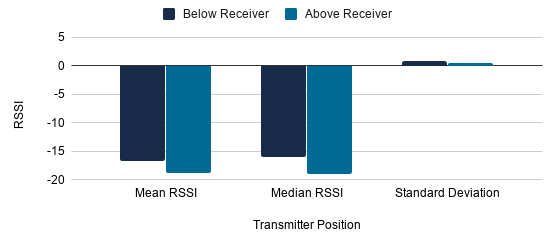}}
\caption{RSSI for transmitter above and below receiver.}
\label{rssi_tx_position}
\end{center}
\vspace{-1mm}
\end{figure}

In addition to power and frequency considerations when calculating RSSI on the receiver, we evaluated if the transmitter position played a role in the stability of RSSI readings (as only one of the two would be placed near the ground). We determined they behaved fairly consistently, with a standard deviation of $0.87$ dBm and $0.60$ dBm when the transmitter was below and above the receiver respectively as seen in Fig.~\ref{rssi_tx_position} .In addition, a mean and median RSSI comparison in Fig.~\ref{rssi_tx_position} demonstrates slight offset in mean and median values indicating no significant difference in transmitter position on RSSI. As a result we opted to keep the transmitter above the receiver.

\section{Experiment and Results}
\subsection{Preliminary Testing}
\begin{figure}[htbp]
\begin{center}
\centerline{\includegraphics[width=0.48\textwidth]{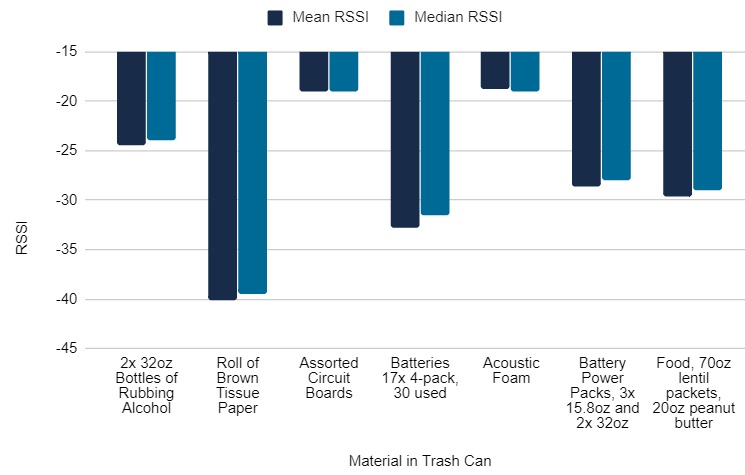}}
\caption{Mean and Median RSSI comparison across different materials.}
\label{rssi_mean_median}
\end{center}
\vspace{-1mm}
\end{figure}

To demonstrate the feasibility of our approach, we needed to ensure the RSSI values generated by the receiver when different materials were placed between the transmitter and receiver would vary. We placed various materials between the transmitter and receiver using a small 2’6” trash can. The resulting mean and median RSSI readings for the variety of materials are depicted in Fig.~\ref{rssi_mean_median}. Based on the observed readings, the mean/median RSSI values were different across different materials. This confirms our hypothesis that RSSI can act as an estimate of material attenuation.

\begin{figure}[htbp]
\begin{center}
\centerline{\includegraphics[width=0.48\textwidth]{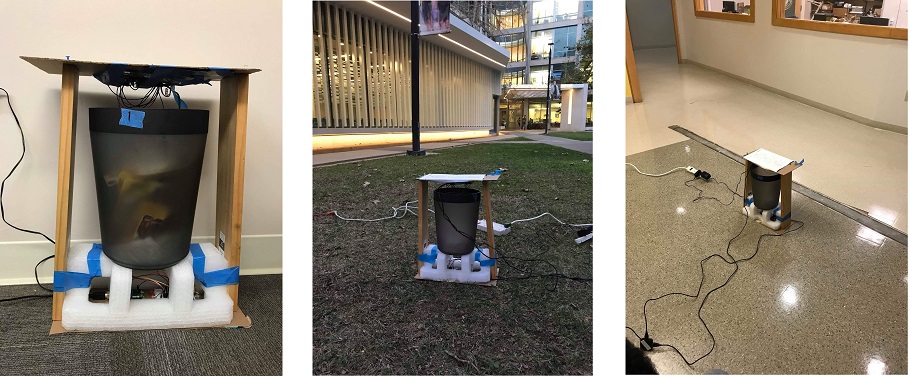}}
\caption{Test environments.}
\label{test_setup}
\end{center}
\vspace{-1mm}
\end{figure}

After establishing that RSSI would change depending on the material placed between the transmitter and receiver, we evaluated the behavior of the system across different environments in which a trash bin may be present (indoor lab against a wall, indoor in an open space, and outdoors). Our test setup in the three aforementioned environments are displayed in Fig.~\ref{test_setup}. We used a test harness to prevent variation in data as a result of variation in our test setup transceiver positions.

\begin{figure}[htbp]
\begin{center}
\centerline{\includegraphics[width=0.48\textwidth]{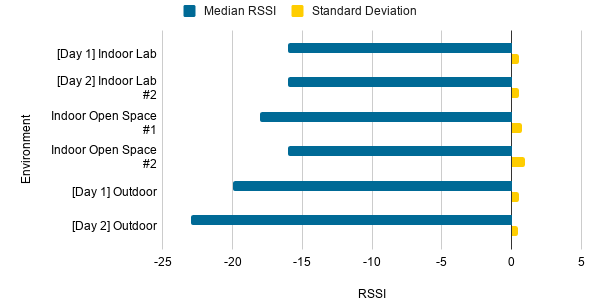}}
\caption{Comparison of Empty Trash Readings.}
\label{rssi_empty_trash}
\end{center}
\vspace{-1mm}
\end{figure}

\begin{figure}[htbp]
\begin{center}
\centerline{\includegraphics[width=0.48\textwidth]{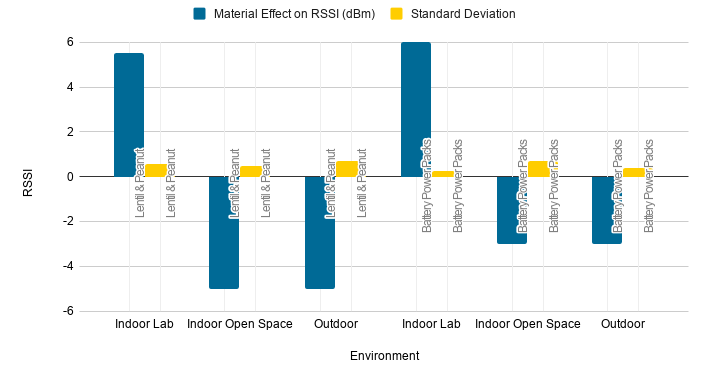}}
\caption{RSSI comparison across environments.}
\label{rssi_envt_comparison}
\end{center}
\vspace{-1mm}
\end{figure}

The environmental testing with just an empty trash bin revealed in Fig.~\ref{rssi_empty_trash} that the outdoor space had the lowest RSSI reading, followed by the indoor open space and then the indoor lab space. The difference in results is likely the result of RF multipath propagation, where radio signals reach the receiver from multiple paths as a result of reflections on nearby objects. The outdoor and open space areas had less nearby objects for the transmitting signal to reflect on, resulting in the lower RSSI readings. The more confined space with nearby walls found in the indoor lab environment yielded the stronger signal likely due to increased reflections on the nearby wall resulting in multipath signal reception. Because the standard deviation in the empty trash bin case from Fig.~\ref{rssi_empty_trash} across the environments was fairly stable, it demonstrated our RSSI approach would likely behave stably with the same accuracy across all environments.
Testing with two materials (food - 90 ounces and battery packs - 110 ounces), we observed interesting results as detailed in Fig.~\ref{rssi_envt_comparison}. Material effect is calculated by subtracting the median RSSI of the empty measurement from the median RSSI with the material present inside the trash bin. Unexpectedly, the RSSI increased in the indoor lab setup case (indicating the signal actually got stronger). Predictably, the indoor open space and the outdoor material effects were the same, likely due to similar relative multipath effects in both cases.

\begin{figure}[htbp]
\begin{center}
\centerline{\includegraphics[width=0.48\textwidth]{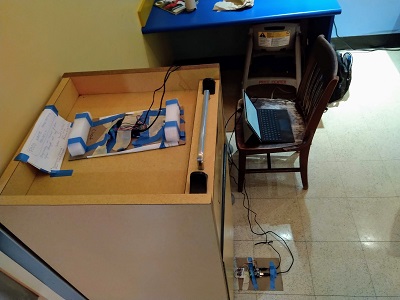}}
\caption{Test setup at Price Center, UC San Diego.}
\label{test_price_center}
\end{center}
\vspace{-1mm}
\end{figure}
\begin{figure}[htbp]
\begin{center}
\centerline{\includegraphics[width=0.48\textwidth]{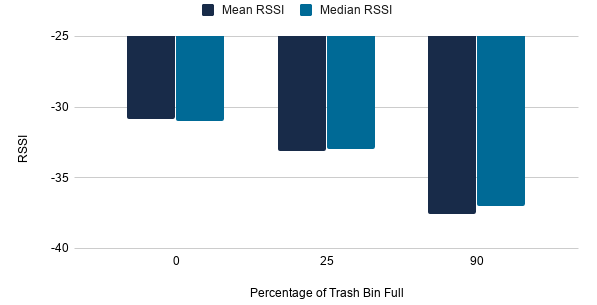}}
\caption{RSSI versus trash level at Price Center, UC San Diego.}
\label{rssi_price_center}
\end{center}
\vspace{-1mm}
\end{figure}
\begin{figure}[htbp]
\begin{center}
\centerline{\includegraphics[width=0.48\textwidth]{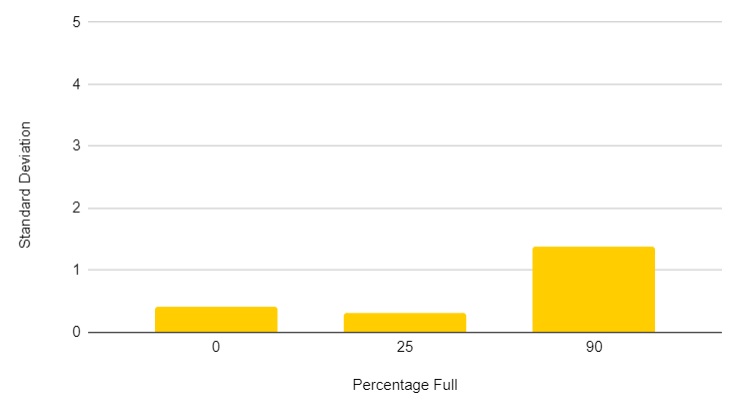}}
\caption{RSSI Standard Deviation at Price Center, UC San Diego}
\label{std_dev_price_center}
\end{center}
\vspace{-1mm}
\end{figure}

\subsection{Live Testing}
Our first set of test data outside of the test harness was taken at the UCSD Price Center as seen in Fig.~\ref{test_price_center}. There we observed as expected the RSSI values dropped from a median reading of -31 to -37 as the garbage level rose from 0 to 90\% as evident in Fig.~\ref{rssi_price_center}. We note in Fig.~\ref{std_dev_price_center}, the standard deviation across readings remain low. This confirms that in a live uncontrolled environment, the RSSI values are stable. While exact metrics on the weight of food present were not attainable, the majority content was visually confirmed to be food due to the trash bin being located in the dining hall. The trend of RSSI decreasing as the waste level increased was evidence towards our hypothesis that correlation between the RSSI and food waste would exist in a real-world scenario. This provided evidence that we would be able to model such a correlation for known waste quantities and provide an estimate of the waste present.

\begin{figure}[htbp]
\begin{center}
\centerline{\includegraphics[width=0.48\textwidth]{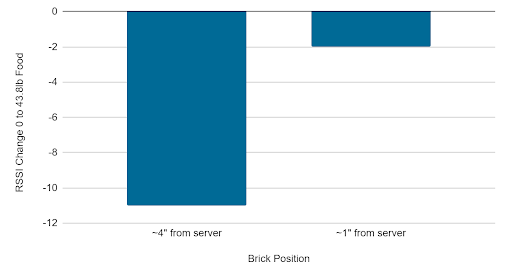}}
\caption{RSSI variation due to distance between bricks.}
\label{brick_distance}
\end{center}
\vspace{-1mm}
\end{figure}

\begin{figure}[htbp]
\begin{center}
\centerline{\includegraphics[width=0.48\textwidth]{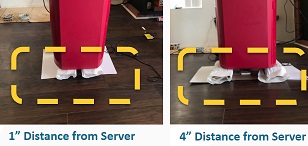}}
\caption{Near and far placement of bricks.}
\label{brick_position}
\end{center}
\vspace{-1mm}
\end{figure}

\begin{figure}[htbp]
\begin{center}
\centerline{\includegraphics[width= 0.4\textwidth, height = 8 cm, keepaspectratio]{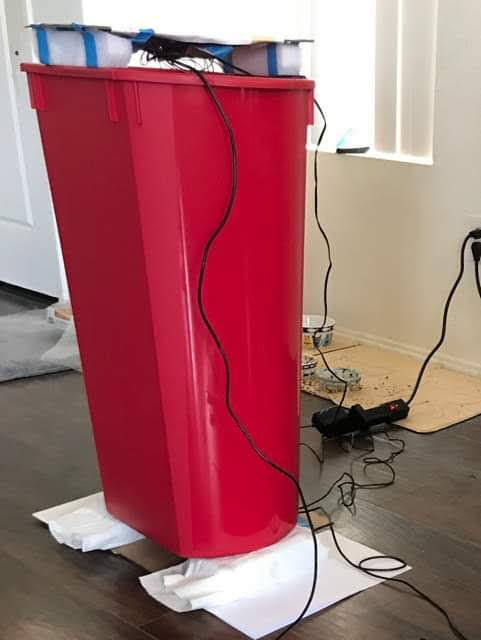}}
\caption{Test setup for food waste estimation.}
\label{test_grocery}
\end{center}
\vspace{-1mm}
\end{figure}

Our second set of data was gathered using fruits and vegetables thrown out by a local grocery story. For this data set, we could control the weight of the food waste in our own trash bin as seen in Fig.~\ref{test_grocery}. We segmented the food waste into five bags weighing 17, 13.8, 13, and 10.6 pounds. While recording RSSI results across different waste levels, we noticed that the position of bricks used to raise the trash bin above one of our RF transceivers shown in Fig.~\ref{brick_position} affected the resulting RSSI values. The effects are captured in Fig.~\ref{brick_distance}. Therefore, we kept the bricks further apart to observe wider variation in RSSI. We note this phenomenon as a potential limitation to our system, requiring the "ground" transceiver to be surrounded by open space if possible to improve the range of RSSI values recorded as the waste contents change.

\begin{figure}[htbp]
\begin{center}
\centerline{\includegraphics[width=0.48\textwidth]{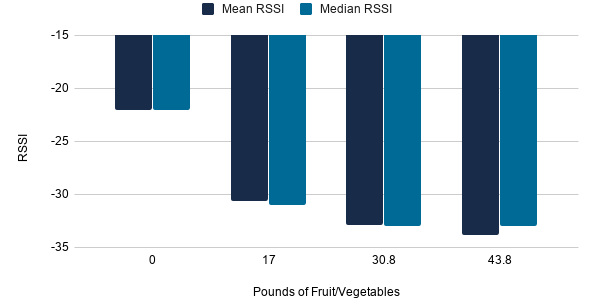}}
\caption{RSSI versus weight of grocery waste.}
\label{rssi_grocery_waste}
\end{center}
\vspace{-1mm}
\end{figure}

\begin{figure}[htbp]
\begin{center}
\centerline{\includegraphics[width=0.48\textwidth]{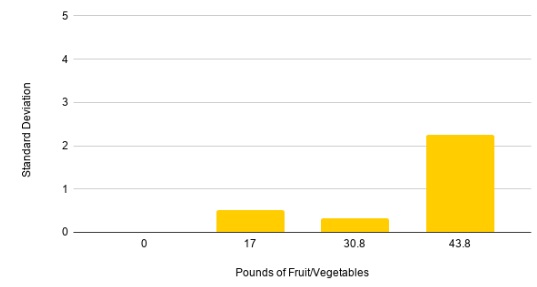}}
\caption{RSSI Standard Deviation for grocery waste}
\label{std_dev_grocery}
\end{center}
\vspace{-1mm}
\end{figure}

We calculated the RSSI values of increasing food waste in our test trash bin using the 17, 13.8 and 13 pound food bags. In Fig.~\ref{rssi_grocery_waste}, we note that the mean RSSI values all increase as expected, however the median RSSI value for the addition of the last 13 pound bag appears to have not caused additional attenuation. We hypothesize this may be due to the proximity of the waste bag to the transmitter antenna, as the RSSI behavior is as expected for the lower weights when the closest waste bag was inches from the antenna. Similar to the Price Center results, standard deviation remained low for the first three weights tested as shown in Fig.~\ref{std_dev_grocery}.

\begin{figure}[htbp]
\begin{center}
\centerline{\includegraphics[width=0.48\textwidth]{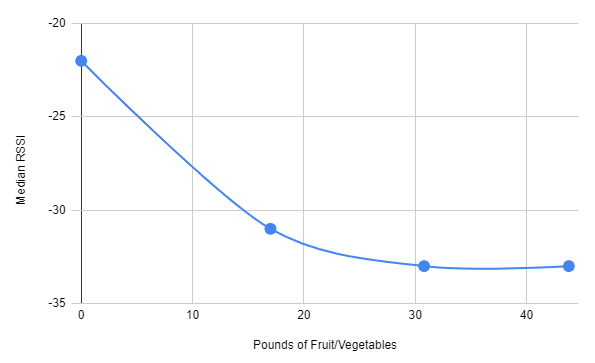}}
\caption{Model for food waste estimation.}
\label{model_food_waste}
\end{center}
\vspace{-1mm}
\end{figure}

Finally, using the data from our previous RSSI grocery waste experiment, we could interpolate the RSSI reading and corresponding food weight in pounds to develop a food waste estimation model for the data set. We then placed our final 10.6 pound food waste bag in the test bin between our transceivers, and recorded a median RSSI value of -27. We could then use the interpolated polynomial from the previous RSSI readings to estimate/predict the food waste amount in pounds, by solving the polynomial for weight:
\begin{equation*}
-0.000161541*x\textsuperscript{3} + 0.0202049*x\textsuperscript{2} - 0.82621*x - 22 = -27
\end{equation*}
\begin{equation*}
    x = 7.2 
\end{equation*}
 Our model predicted 7.2lb, resulting in 32\% error from the actual 10.6lb weight. Fig. ~\ref{model_food_waste} shows a visual approximation of the interpolated estimation model.

\section{Conclusion}
Our initial results are promising in nature, demonstrating with live test results that RSSI decreases as more food waste is present in the trash bin. This correlation between RSSI level and waste level is explained by the additional attenuation the additional waste causes. Another positive observation was the relatively low standard deviation observed across test results, indicating the measurement system would be stable in a non lab environment. A limitation during testing was access to live test data, as dining hall food waste access was limited by bureaucratic and health safety rules, and reclaiming test data from grocery dumpsters poses health hazards. We note in further work, we can potentially achieve a higher fidelity estimation using Software Defined Radios to measure the actual energy received, instead of using RSSI. We could also use a more mono-directional antenna to ensure our signal is passing through the material as expected, improving the accuracy of the resulting attenuation readings. Improvements in hardware will likely improve on results, however realistic deployments with such a system would face financial restrictions favoring the cheaper, coarser grain results achieved by our current system. The insight of using attenuation to infer the material present between the RF transceivers at the core allows us to measure the presence of something without seeing or physically touching it. There may be additional applications in which sensing is limited where such an RF attenuation/RSSI sensing approach could prove beneficial, and further research into such an application could be done.

\section*{Acknowledgment}
We thank professor Aaron Schulman for his feedback, advice, guidance, and lab space. Additionally we thank Jim Duino from Viasat for his insight on RF and Alex Koohmarey for his assistance in attaining the grocery waste data.

\bibliographystyle{IEEEtran}

\begin{thebibliography}{10}
\providecommand{\url}[1]{#1}
\csname url@samestyle\endcsname
\providecommand{\newblock}{\relax}
\providecommand{\bibinfo}[2]{#2}
\providecommand{\BIBentrySTDinterwordspacing}{\spaceskip=0pt\relax}
\providecommand{\BIBentryALTinterwordstretchfactor}{4}
\providecommand{\BIBentryALTinterwordspacing}{\spaceskip=\fontdimen2\font plus
\BIBentryALTinterwordstretchfactor\fontdimen3\font minus
  \fontdimen4\font\relax}
\providecommand{\BIBforeignlanguage}[2]{{%
\expandafter\ifx\csname l@#1\endcsname\relax
\typeout{** WARNING: IEEEtran.bst: No hyphenation pattern has been}%
\typeout{** loaded for the language `#1'. Using the pattern for}%
\typeout{** the default language instead.}%
\else
\language=\csname l@#1\endcsname
\fi
#2}}
\providecommand{\BIBdecl}{\relax}
\BIBdecl

\bibitem{fao_1}
FAO, ``Food loss and food waste,'' Available at
  \url{http://www.fao.org/food-loss-and-food-waste/en/}.

\bibitem{dusseldorfsave}
M.~D{\"u}sseldorf, ``Study conducted for the international congress - save
  food!''

\bibitem{fao_2}
FAO, ``Food wastage footprint \& climate change,'' Available at
  \url{http://www.fao.org/3/a-bb144e.pdf}.

\bibitem{ucsd_zerowaste}
UCSD, ``Zero waste plan as of september 2019,'' Available at
  \url{https://sustainability.ucsd.edu/files/UCSanDiegoZeroWastePlan.pdf}.

\bibitem{ha2018learning}
U.~Ha, Y.~Ma, Z.~Zhong, T.-M. Hsu, and F.~Adib, ``Learning food quality and
  safety from wireless stickers.'' in \emph{HotNets}, 2018, pp. 106--112.

\bibitem{rappaport1996wireless}
T.~S. Rappaport \emph{et~al.}, \emph{Wireless communications: principles and
  practice}.\hskip 1em plus 0.5em minus 0.4em\relax prentice hall PTR New
  Jersey, 1996, vol.~2.

\bibitem{ryan2003radio}
P.~L. Ryan, ``Radio frequency propagation differences through various
  transmissive materials.'' 2003.

\bibitem{nguyen2015wireless}
S.~D. Nguyen, N.~N. Le, P.~T. Lam, E.~Fribourg-Blanc, C.~M. Dang, and
  S.~Tedjini, ``A wireless sensor for food quality detection by uhf rfid
  passive tags,'' in \emph{2015 International Conference on Advanced
  Technologies for Communications (ATC)}.\hskip 1em plus 0.5em minus
  0.4em\relax IEEE, 2015, pp. 258--263.

\bibitem{potyrailo2012battery}
R.~A. Potyrailo, N.~Nagraj, Z.~Tang, F.~J. Mondello, C.~Surman, and W.~Morris,
  ``Battery-free radio frequency identification (rfid) sensors for food quality
  and safety,'' \emph{Journal of agricultural and food chemistry}, vol.~60,
  no.~35, pp. 8535--8543, 2012.

\bibitem{smart_waste_management}
K.~V.~J. Vedant Nitin~Agnihotri, Atul~Srivastava, ``Smart waste management
  system using zig-bee,'' \emph{International Journal for Research in Applied
  Science \& Engineering Technology (IJRASET)}, vol.~5, no.~VI, 2017.

\bibitem{lata2016iot}
K.~Lata and S.~S. Singh, ``Iot based smart waste management system using
  wireless sensor network and embedded linux board,'' \emph{Int. J. Curr.
  Trends Eng. Res}, vol.~2, pp. 210--214, 2016.

\bibitem{cerchecci2018low}
M.~Cerchecci, F.~Luti, A.~Mecocci, S.~Parrino, G.~Peruzzi, and A.~Pozzebon, ``A
  low power iot sensor node architecture for waste management within smart
  cities context,'' \emph{Sensors}, vol.~18, no.~4, p. 1282, 2018.

\bibitem{zou2014radio}
Z.~Zou, Q.~Chen, I.~Uysal, and L.~Zheng, ``Radio frequency identification
  enabled wireless sensing for intelligent food logistics,''
  \emph{Philosophical Transactions of the Royal Society A: Mathematical,
  Physical and Engineering Sciences}, vol. 372, no. 2017, p. 20130313, 2014.

\bibitem{hall2009progressive}
K.~D. Hall, J.~Guo, M.~Dore, and C.~C. Chow, ``The progressive increase of food
  waste in america and its environmental impact,'' \emph{PloS one}, vol.~4,
  no.~11, p. e7940, 2009.

\bibitem{benkic2008using}
K.~Benkic, M.~Malajner, P.~Planinsic, and Z.~Cucej, ``Using rssi value for
  distance estimation in wireless sensor networks based on zigbee,'' in
  \emph{2008 15th International Conference on Systems, Signals and Image
  Processing}.\hskip 1em plus 0.5em minus 0.4em\relax IEEE, 2008, pp. 303--306.

\bibitem{electronic2016rfm22b}
H.~Electronic, ``Rfm22b/23b ism transceiver module,'' \emph{RFM22B/23B
  datasheet}, 2016.

\bibitem{mccauley2014radiohead}
M.~McCauley, ``Radiohead packet radio library for embedded microprocessors,''
  \emph{Avaliable online: http://www. airspayce.
  com/mikem/arduino/RadioHead/(accessed on 20 November 2018)}, 2014.

\bibitem{rf_cafe}
I.~MECA~Electronics, ``Rf engineering calculators,'' Available at
  \url{http://www.rfcafe.com/references/calculators/calculator-list.htm}.

\end{thebibliography}

\end{document}